# Low Temperature MOCVD Synthesis of high-mobility 2D InSe

Robin Günkel[1,2], Oliver Maßmeyer[1,2], Markus Stein[3], Sebastian Anhäuser[1,2], Kalle Bräumer[3], Rodrigo Sandoval Rodriguez[3], Daniel Anders[3], Badrosadat Ojaghi Dogahe[1,2], Max Bergmann[1,2], Milan Solanki[1,2], Nils Fritjof Langlotz[1,2], Johannes Glowatzki[1,2], Jürgen Belz[1,2], Andreas Beyer[1,2], Gregor Witte[1,2], Sangam Chatterjee[3], Kerstin Volz[1,2, *],

[1]*mar.quest | Marburg Center for Quantum Materials and Sustainable Technologies, Philipps-University Marburg, Germany*
[2]*Department of Physics, Philipps-University Marburg, Germany*
[3]*Institute of Experimental Physics I and Center for Materials Research, Justus Liebig University Giessen, Germany*

* corresponding author: kerstin.volz@physik.uni-marburg.de

ABSTRACT
Two-dimensional (2D) indium selenide (InSe) is a layered semiconductor with high electron mobility and a tunable band gap ranging from 1.25 eV in the bulk to 2.8 eV in the monolayer limit. These properties make these materials strong candidates for future logic and optoelectronic devices. However, growing phase-pure InSe remains challenging due to the complex indium–selenium (In–Se) phase diagram. This complexity and the sensitivity of chemical precursors to growth conditions make it difficult to control which In–Se phase forms during synthesis during, e.g., metal-organic chemical vapor deposition (MOCVD).
Despite the challenges, MOCVD is considered the most promising approach for growing InSe, as it enables wafer-scale, uniform, and controllable deposition—key requirements for device integration. In this study, we present a systematic investigation of InSe synthesis via MOCVD on c-plane sapphire substrates at low temperatures, which are highly relevant for various integration schemes. By varying the Se/In precursor ratio and the growth temperature, we create a phase diagram that covers the In-rich, equal stoichiometric, and Se-rich $In_xSe_y$ phases. Raman spectroscopy and atomic force microscopy, supported by energy dispersive X-ray spectroscopy and scanning transmission electron microscopy, confirm conditions, under which the formation of 2D InSe is observed. Atomically-resolved cross-sectional scanning transmission electron microscopy also reveals an epitaxial alignment of the InSe with the sapphire substrate mediated by a specific interface reconstruction. The epitaxial alignment is verified by in-plane X-ray diffraction across large length scales. Samples grown under optimized conditions exhibit a strong optical absorption in the visible range and especially a comparably high electron mobility underlining the potential of the MOCVD-grown material for future applications.

INTRODUCTION
The study of two-dimensional (2D) materials has attracted considerable interest, particularly following the exfoliation of graphene and the studies of its electronic structure that led to the 2010 Nobel Prize in Physics[1,2]. Graphene, along with other 2D materials, offers significant potential for further miniaturization of logic-based devices[3–5]. However, for these materials to be suitable for future applications, they must not only exhibit superior properties compared to existing solutions, but also enable a scalable and cost-efficient synthesis, compatible with current technology platforms[6].

Among 2D materials, indium selenide (InSe) stands out for its high field effect mobility, making it a promising candidate for next-generation logic devices[7,8]. Prototype devices based on InSe have already been demonstrated[9–12]. Furthermore, light matter interaction[13–17] as for example the strongly layer-tunable band gap from 1.25 eV in the bulk to 2.8 eV in the monolayer limit [12,17–19] and potential optoelectronic applications[13] as for example photodetectors[20–22] or gas sensors[23,24] based on InSe are investigated. For practical applications InSe must be synthesized using scalable methods that ensure reproducibility[25,26]. One challenge for the bottom-up synthesis using different growth techniques is the large number of existing In-Se phases[27,28]. Nevertheless, first thin film deposition experiments demonstrate phase control for example via Molecular Beam Epitaxy (MBE)[29–34], Atomic Layer Deposition (ALD)[35], Physical Vapor Deposition (PVD)[36,37] or Pulsed Laser Deposition (PLD)[38,39]. Furthermore, Chemical Vapor Deposition (CVD) of InSe has previously been demonstrated on mica substrates at elevated temperatures of 600 °C[40] and 630 °C[41]. On c-plane sapphire, metal-organic chemical vapor deposition (MOCVD) of InSe using trimethylindium (TMIn) and diethyl selenium (DESe) has been reported within a process window of 500 °C to 600 °C[42]. However, for compatibility with complementary metal-oxide-semiconductor (CMOS) technology, reduced thermal loads of ≤ 400 °C are required to enable in-line integration with back-end-of-line (BEOL) processes[43–49]. Below this threshold temperature, single-source precursor approaches have been explored for $In_xSe_y$ deposition[50,51]; however, these have yet to achieve the formation of layered 2D InSe. Recently, a modulated precursor supply strategy has enabled the growth of phase-stable, layered InSe in the 350–500 °C temperature range[11] with smooth 2D layers demonstrated at temperatures of 500 °C and above. This first report on 2D InSe synthesis by MOCVD demonstrates its growth in a vertical showerhead reactor using TMIn and dimethyl selenide (DMSe) as precursor sources[11]. To reduce the deposition temperature of smooth layers further, Se-precursors featuring a lower decomposition temperature compared to DMSe are important, as the effectiveness of MOCVD growth depends strongly on the activation energy for the thermal decomposition of the employed precursors, which determines the lower limit of their efficient use. Among the common organo-selenium precursors — DMSe, DESe, and diisopropyl selenide (DiPSe) — DiPSe exhibits the lowest activation energy for thermal decomposition[52], making it particularly suitable for low-temperature MOCVD processes. DiPSe has been successfully used for the MOCVD growth of several other chalcogenide materials, including $Cu(GaIn)Se_2$[53], 2D GaSe on Si(111)[54] and 2D $WSe_2$[55,56], further underlining its versatility.

In this work, we focus on the controlled growth and comprehensive characterization of 2D InSe synthesized via MOCVD. While previous studies have successfully demonstrated the MOCVD synthesis of $β-In_2Se_3$[11,57] and even realized ferroelectric transistors based on this phase[58], these efforts did not address the growth challenges or application potential of equal stoichiometric InSe. We go beyond these reports by demonstrating the phase-selective and low-temperature growth of 2D InSe films using TMIn and DiPSe – precursors well-established in the epitaxy of III–V and II–VI semiconductors[59]. Our goal is to enable controlled low-temperature growth of 2D InSe films compatible with CMOS requirements. Moreover, we aim at demonstrating the superior, application-relevant properties of the 2D phase of InSe grown by MOCVD in this temperature window.

In this work, we first demonstrate, how temperature and precursor ratio influence the formation of distinct $In_xSe_y$ phases, as identified by Raman spectroscopy. Atomic force microscopy (AFM) reveals the transition from isolated flakes to continuous layered films with increasing growth time. Subsequently, scanning transmission electron microscopy (STEM) and in-plane X-ray diffraction (XRD) confirm the epitaxial alignment of the InSe layers with the c-plane sapphire substrate. Optical transmission measurements reveal well-defined excitonic absorption features, while optical-pump terahertz-probe (OPTP) spectroscopy uncovers ultrafast photoconductivity dynamics and demonstrates a high carrier mobility in continuous InSe films.

RESULTS and DISCUSSION

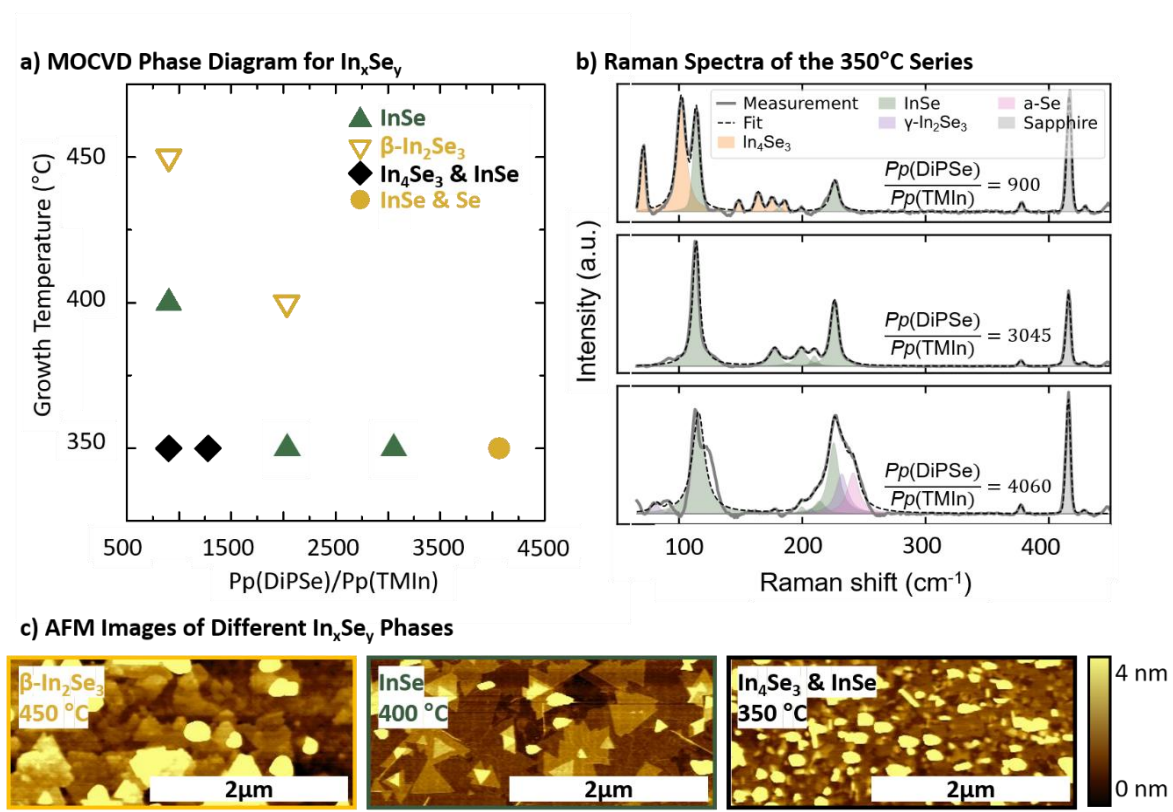

Figure 1: a) Phase diagram showing the dominant $In_xSe_y$ phases as a function of the growth temperature (350 °C to 450 °C) and the precursor ratio (900 to 4060). The targeted InSe is depicted as green triangles; Se-rich phases appear in yellow, and In-rich phases in black. b) Exemplary Raman spectra used to identify the different $In_xSe_y$ phases obtained at 350 °C reactor temperature and a varied precursor ratio of Pp(DiPSe)/Pp(TMIn) = 900 up to 4060 (from top to bottom). c) AFM images illustrating the surface morphology of representative $In_xSe_y$ samples grown at a constant precursor ratio of 900 and temperatures of 450 °C (yellow frame), 400 °C (green frame), and 350 °C (black frame), respectively.

Figure 1a presents the MOCVD phase diagram of $In_xSe_y$ as a function of the precursor ratio Pp(DiPSe)/Pp(TMIn) at constant TMIn supply and reactor temperature of 350 °C, 400°C and 450°C, respectively. Raman spectroscopy was used to identify the resulting phases by matching the measured spectra with reference data reported for for InSe[34], $β-In_2Se_3$[34], $γ-In_2Se_3$[34], amorphous Selenide[60] and $In_4Se_3$[61]. The corresponding AFM and Raman measurements for all samples are provided in Supporting Information Figures SI1 and SI2, respectively. As both, the growth temperature and the selenium supply increase, the material transitions from In-rich phases (e.g., $In_4Se_3$) to Se-rich phases (e.g., $β-In_2Se_3$ or $γ-In_2Se_3$), with phase-pure InSe forming under intermediate conditions. Figure 1b illustrates this effect in more detail by varying the DiPSe offer at a constant TMIn flux and at a constant temperature, resulting in distinct $In_xSe_y$ phases, each identified by their characteristic Raman signatures as referenced above. The emergence of specific phases is highly sensitive to the Se-to-In ratio. Notably, this study also demonstrates that phase-pure InSe can be synthesized at temperatures as low as 350 °C, as shown by the central spectrum of Figure 1b. Furthermore, the sample´s surface morphology correlates with the respective $In_xSe_y$ phase: the AFM analysis (Figure 1c) reveals that InSe forms well-defined, layered triangular flakes approximately 1 µm in size, whereas mixed-phase samples or samples mainly consisting of other $In_xSe_y$ phases show rougher surface morphologies. Before we now turn to investigate the InSe sample in more detail, we would like to emphasize that the nucleation conditions for a 2D-material are very critical and hence, defined conditions need to be adjusted during the growth of the first layer. Figure SI3 exemplarily shows the large influence of an Indium pre-deposition step prior to In-Se growth. As In will be also deposited from the reactor walls in the cleaning step of the substrate, we have consequently included an In pre-deposition layer in all samples shown in the following, to have comparable starting conditions.

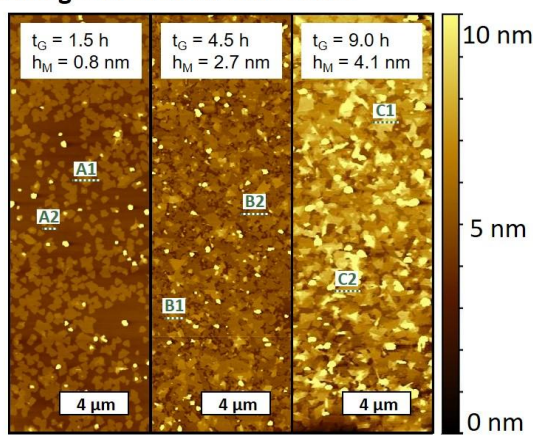
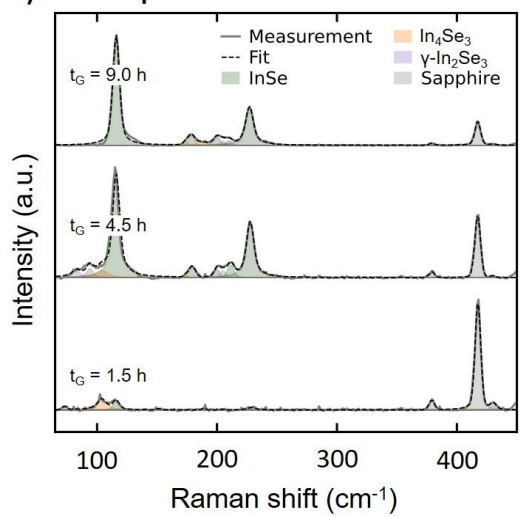
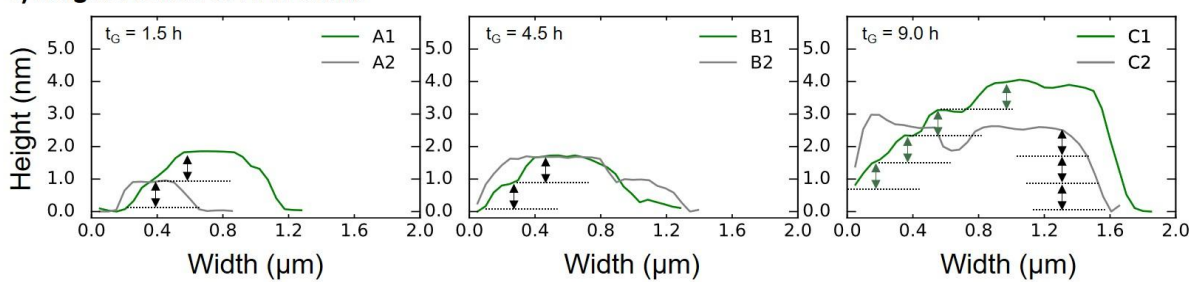

Figure 2: a) Atomic force microscopy images of InSe samples grown for different durations: $t_G$ = 1.0 h, 4.5 h, and 9.0 h, corresponding to median film heights of $h_M$ = 0.8 nm, 2.7 nm, and 4.1 nm, respectively. b) Raman spectra showing the emergence of characteristic InSe vibrational modes, most pronounced in the 9 h sample (top box) and clearly visible in the 4.5 h sample (middle box), confirming the formation of the InSe phase. Compared to the sapphire background signal, the intensity of the InSe Raman modes increases significantly with longer growth duration, as seen from the 1.5 h sample (bottom box) to the 9 h sample. c) Three representative height profiles, from left to right, corresponding to samples grown for 1.5 h, 4.5 h, and 9 h, respectively.

Figure 2a depicts an AFM height series of samples grown at 400 °C with the optimized Se/In ratio to achieve single-phase InSe, but for different times to investigate the morphology when the thickness is changing. As the growth time increases from 1.5 hours to 9.0 hours (from left to right), the median height $h_M$ of the height distribution in the AFM images concomitantly increases from 0.8 nanometers to 4.1 nanometers, corresponding to an increase from approximately one to five InSe layers, respectively. Furthermore, the line scans in Figure 2c reveal that the height profiles consist of multiples of approximately 0.8 nm, consistent with the monolayer thickness of InSe. Figure 2b displays the averaged and subsequently fitted Raman spectra corresponding to the samples. These fits consistently indicate the presence of the targeted InSe phase. However, the spectrum of the 1.5-hour sample shows additional Raman modes associated with $In_4Se_3$[61], and the peak fit for the 4.5-hour sample includes features attributed to γ-$In_2Se_3$[34]. In contrast, the 9-hour sample exhibits a distinct Raman signature characteristic of phase-pure InSe. The presence of secondary phases in the thinner films may serve as an initial indication of a covalently bonded interfacial layer on the sapphire surface and/or an incomplete nucleation process during the early stages of growth. We discuss this in context with the electron microscopy data of the interface between the substrate and the InSe (Figure 3). Calculating the difference between the A′$_1$(1) and A′$_1$(2) Raman mode positions of InSe, as motivated by Molas et al. [62] reveals a decrease in difference with increasing growth time, i.e., increasing layer thickness. The difference in position decreases from 113 cm$^{-1}$ to 112 cm$^{-1}$ and finally to 111 cm$^{-1}$ when increasing the InSe thickness from approximately 2 layers (expected height ~1.6 nm) via 3–5 layers (expected height ~2.4–4.0 nm) to 5–7 layers (expected height ~4.0–5.6 nm), respectively. Our previously determined median heights of the three samples depicted in Figure 2 lie within these thickness ranges and also show the position of the Raman modes expected from literature, confirming the consistency between morphological and spectroscopic characterization. In addition, the AFM images (Figure SI4) of the 9-hour grown sample across the wafer underline the homogeneous synthesis of InSe at wafer-scale.

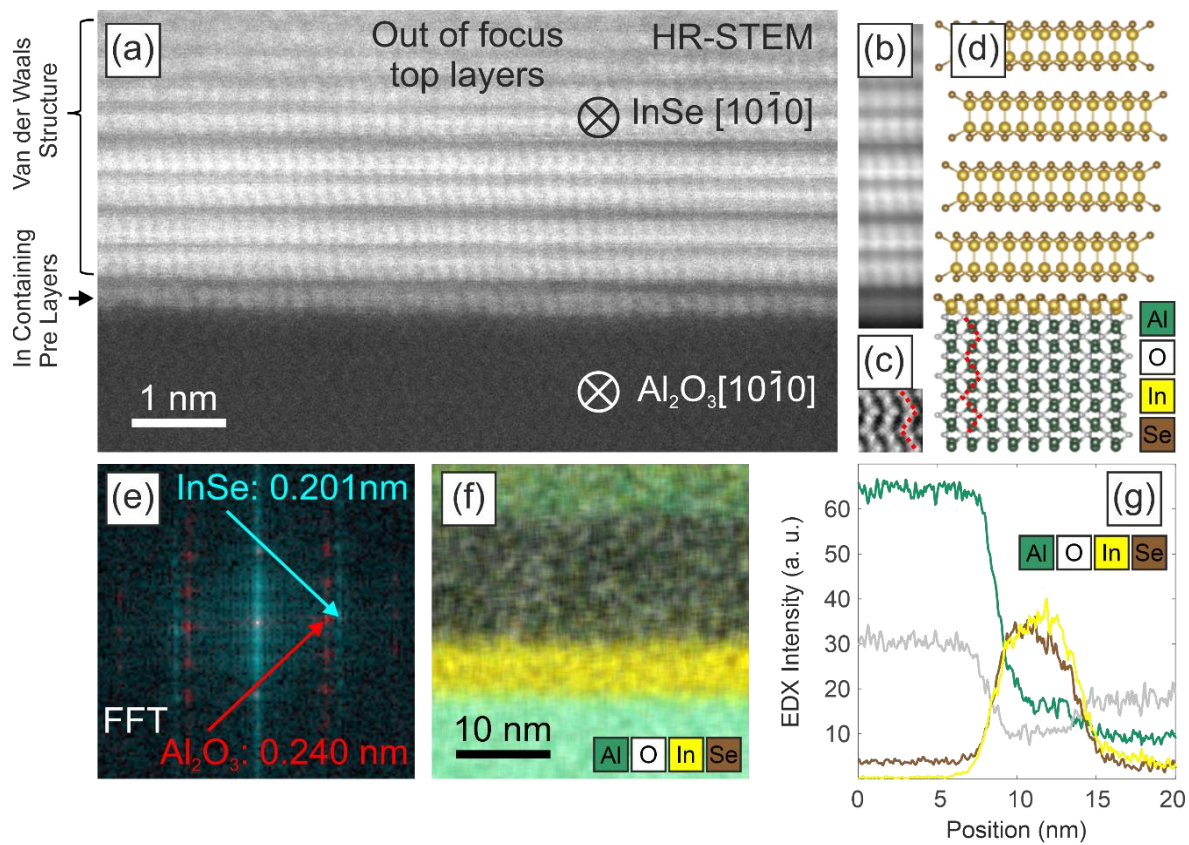

Figure 3: a) Cross-sectional scanning transmission electron microscopy (STEM) image of InSe (9 hours growth time). b) and c) Position averaged high-resolution STEM image of InSe and sapphire, respectively. d) Ball-and-stick model for the InSe van-der-Waals-like structure on sapphire. e) FFT of a) showing the periodicity of Al$_2$O$_3$ {11-20} planes (red) and the InSe {11-20} planes (cyan). f) EDX map of the InSe cross section showing the location of aluminum (green), oxygen (white), indium (yellow) and selenium (brown) using the Al-Kα, O-Kα, In-Lα, Se-Kα lines, respectively. g) Line profile extracted from the EDX maps.

To further investigate the epitaxial alignment and interfacial structure between the InSe and the sapphire substrate, high-resolution scanning transmission electron microscopy (HR-STEM) is performed on cross-sectional samples, as illustrated in Figure 3. The high-resolution STEM cross-section image (Figure 3a) confirms the identified 2-dimensional γ-InSe phase. Due to an increased signal-to-noise ratio due to the low electron dose that could be used on these samples, the crystal structure becomes more evident in the position averaged images of the InSe layers (Figure 3b) and the substrate region (Figure 3c). A ball-and-stick model depicting the arrangement of the hetero interface is depicted in Figure 3d. The crystallographic alignment between the sapphire substrate and the InSe layers is clearly visible from the FFT (Figure 3e). Here, the {11-20} spots of the InSe are parallel to the {11-20} spots of the sapphire. The lattice spacings are determined to be 0.201 nm and 0.240 nm for InSe and $Al_2O_3$, respectively. Both values are consistent with the expected lattice spacings of $d_{γ-InSe}^{\{11-20\}}$ = 0.201 nm and $d_{Al2O3}^{\{11-20\}}$ = 0.240 nm, underlining unstrained growth of the van-der-Waals material on the substrate.

Next, we investigate the interface between the sapphire and the InSe in more detail: the c-plane of pure sapphire is expected to be either aluminum or oxygen terminated.[63] However, the higher intensity observed in the HAADF image at the interface between the first Se layer of the 2D InSe and the last oxygen layer of sapphire substrate (Figure 3a, 3b) suggests the presence of an initial nucleation layer containing In and/or Se, probably covalently bonded to the substrate. This is expected from the TMIn seeding approach and the following Se supply outlined in Figure SI3. In-containing interlayers have also been observed in MBE experiments[30]. An analogue study on GaS grown on sapphire reported a comparable Ga layer bound to the surface [64]. To further confirm the 1:1 stoichiometry of the InSe layers, complementary energy dispersive X-ray spectroscopy (EDX) measurements were carried out. A map of the hetero interface is shown in Figure 3f. The individual concentration profiles derived from this map are shown in Figure 3g. Indeed, an equal concentration of In and Se is found within the InSe layers reflecting the 1:1 stoichiometry of the γ-InSe phase. This interfacial $In_xSe_y$ layer, which we observe at the interface might well explain the small Raman peaks of non-InSe-phases for very thin layers concluded from Figure 2. It should also be noted that due to re-absorption of the low-energy X-rays within the comparably thick TEM sample, the Al and O profiles can only be interpreted qualitatively.

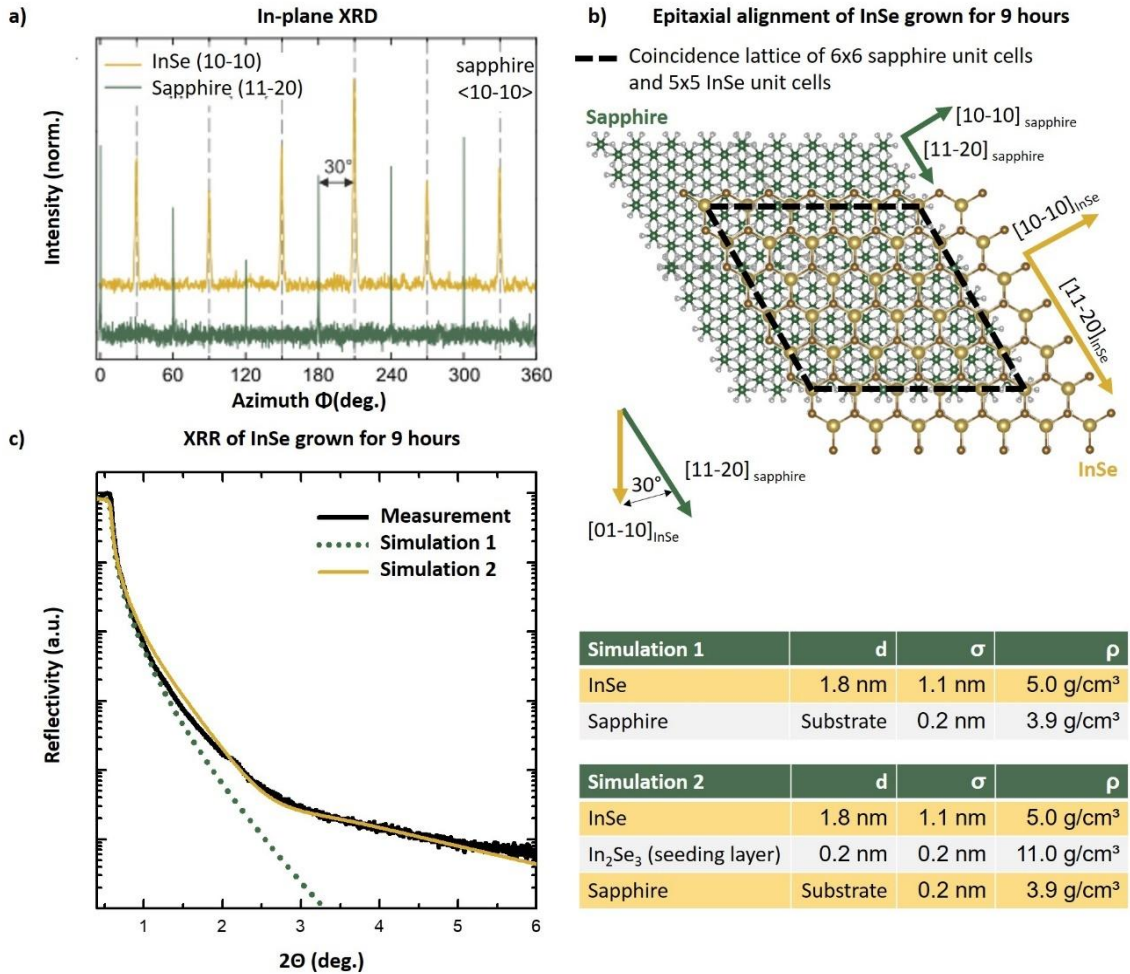

Figure 4: a) In-plane XRD φ-scan of {11-20} reflections of sapphire (green) and of {10-10} reflections of InSe (yellow), respectively. The corresponding azimuthal alignment of InSe to sapphire is shown by a top view model of the lattice planes in b). The coincidence lattice is shown in black, where 6 × 6 sapphire unit cells align with 5 × 5 InSe unit cells. c) X-ray reflection (XRR) with simulation 1 (InSe: 1.8 nm, σ = 1.1 nm, ρ = 5 g/cm³; sapphire: σ = 0.2 nm, ρ = 3.9 g/cm³) and simulation 2 with the additional $In_2Se_3$ nucleation layer (InSe: 1.8 nm, σ = 1.1 nm, ρ = 5 g/cm³; $In_2Se_3$ layer: 0.2 nm, σ = 0.2 nm, ρ = 11.0 g/cm³; sapphire: σ = 0.2 nm, ρ = 3.9 g/cm³).

To verify that the azimuthal InSe orientation relative to the sapphire substrate is the same at the wafer scale as initially observed in the local STEM cross-section analysis, we perform in-plane XRD measurements of the 9-hour grown sample (Figure 4a). The angle between the sapphire {11-20} plane and the InSe {10-10} plane is determined to be 30 degrees, meaning that the [10-10] vectors of both materials are parallel. These findings confirm that the InSe, which exhibited a local alignment to the sapphire substrate as revealed by STEM analysis, maintains a consistent epitaxial relationship with the $Al_2O_3$ surface across the wafer. This alignment is illustrated by the schematic model in Figure 4b and corresponds to a calculated coincidence lattice, where 6 × 6 sapphire unit cells align with 5 × 5 InSe unit cells. X-ray reflectivity (XRR) analysis (Figure 4c) is performed to verify the total thickness of the layered InSe crystal as well as to confirm the interfacial $In_xSe_y$ layer, as a structural continuation of the sapphire substrate, on a larger lateral scale than accessible by STEM. As a result of the STEM investigations (Figure 3f) two simulations, one without an $In_xSe_y$ interface layer on the sapphire (simulation 1), and a second one with an $In_xSe_y$ interface on the sapphire (simulation 2), were performed using X-ray Calc3[65]. Simulation 2 confirms the presence of an In-containing interface on the sapphire that forms a layer between the sapphire substrate and InSe, as seen in the STEM cross-sectional image supporting the expected growth process. As shown in the AFM image (Figure 2a), the amount of InSe layers locally varies by about 4 layers, resulting in lower material density on average for thicker layers, which causes a rough surface effect in the XRR signal. Furthermore, the simulated InSe layer thickness of 1.8 nm differs from the median height of 4.1 nm determined from the corresponding AFM images. This can be partially explained by the XRR simulated surface roughness of 1.1 nm. Finally, the XRR measurement supported by simulation 2 shows that there is an $In_xSe_y$ interlayer between the sapphire substrate and the 2D InSe on a wafer scale, as expected from the growth process (TMIn seeding) and already observed by STEM EDX and HAADF analysis (Figure 3).

To assess the material's technological applicability, we perform a comprehensive functional characterization of the InSe films as shown in Figure 5. Optical transmission measurements provide information relevant for photonic applications. Finally, terahertz (THz) time-domain spectroscopy following optical excitation enables the extraction of charge carrier mobility, a critical figure of merit for electronic device performance.

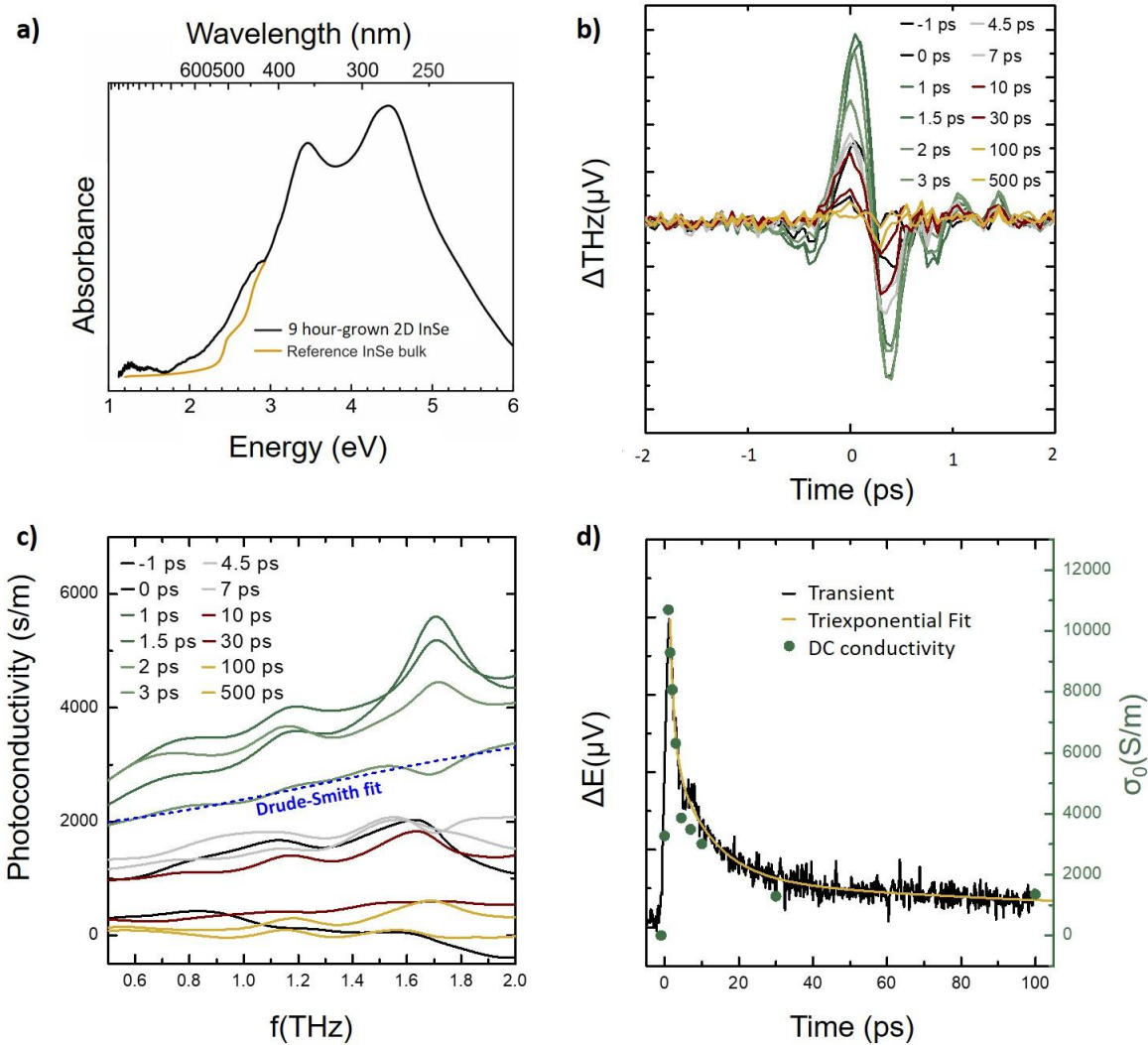

Figure 5: a) Optical absorption spectrum of InSe grown for 9 hours on c-plane double side polished sapphire at growth temperature of 400 °C measured in transmission geometry (black). Reference measurement (yellow) of bulk InSe adapted from Segura et al.[66]. b). Differential THz time-domain traces obtained by subtracting the THz waveform of the optically excited sample from that of the unexcited sample, shown for various temporal delays between optical-pump and THz-probe. c) Real part of the extracted photoconductivity in the THz frequency range for the different temporal delays, together with an exemplary fit of the Drude–Smith model. d) Transient signal at the peak of the differential THz trace (black line), alongside the extracted DC conductivity from the Drude–Smith fits (green spheres). A tri-exponential fit (yellow line) is used to determine the decay dynamics of the light-induced photoconductivity.

To investigate the optical properties of MOCVD-grown InSe, absorption measurements are conducted in the range of 1 eV to 6 eV using a transmission geometry. For these measurements, the InSe sample was grown on double-side-polished c-plane sapphire with a layer number or thickness similar to that of the 9-hour growth sample (as shown in AFM data in Figure 2a and Figure SI4). Its optical absorption spectrum is shown in Figure 5a in black, which is compared with a reference bulk measurement by Segura et al. [66] (in yellow). For comparability, the reference data for an estimated 50 nm thick InSe layer is extracted and fitted to the intensity of our measurement at 1.2 eV. The peaks at 2.9 eV, 3.2 eV and 4.2 eV are consistent with *ab initio* calculations that combine many-body perturbation theory methods, such as the GW method, and the Bethe-Salpeter equation[67]. It can be seen that the three peaks correspond to calculated excitations in the plane of the InSe, which fits very well with the geometrical conditions of the transmission experiment, since the light here is incident parallel to the normal vector (in the c direction). Consequently, the InSe exhibits the expected optical absorption properties, which pave the way for device applications MOCVD grown InSe.

Once the InSe flakes coalesce into a continuous layer, OPTP measurements reveal a pronounced transient photoconductivity response. The strong photoinduced change in the transmitted THz field confirms the generation of mobile charge carriers upon optical excitation, with the signal evolving systematically as a function of pump–probe delay (Figure 5b). The frequency-dependent photoconductivity spectra are analyzed using the Drude–Smith model, which accounts for partially localized carrier transport due to scattering and back reflection (Figure 5c). This analysis yields an effective electron mobility of about·1200 cm²/Vs assuming an effective electron mass of 0.1 times the mass of a free electron[68]. This value is in excellent agreement with reported mobilities for bulk InSe[7], suggesting that the favorable transport properties of the material are largely preserved during the MOCVD growth of few-layer thick films. The temporal evolution of the photoconductivity follows a multi-exponential decay after excitation (cf. Figure 5d). An initial ultrafast decay with a time constant of about 1ps is followed by a slower component on the order of 10ps. A residual photoconductivity persists with a much slower decay time in the range of 200ps, likely due to recombination from trap states or defects and is subject of further investigations.

CONCLUSION

This study demonstrates the successful synthesis of phase-pure films of the 2D material InSe on c-plane sapphire substrates using a horizontal metal-organic chemical vapor deposition (MOCVD) system. Trimethylindium (TMIn) and diisopropyl selenide (DiPSe) serve as precursors at technologically relevant growth temperatures between 350 and 450 °C. A pre-deposition of TMIn is employed to establish a reproducible indium-based seeding layer. By optimizing the DiPSe-to-TMIn precursor ratio, phase purity is achieved, resulting in the exclusive formation of InSe at temperatures as low as 350 – 400 °C. Raman spectroscopy confirms – for optimized growth conditions - the InSe phase, with characteristic vibrational modes indicative of few-layer, high-quality material. Growth time series studied via atomic force microscopy (AFM) show the evolution from isolated triangular InSe flakes to continuous multilayer films. The epitaxial alignment and high crystallinity of the InSe films on sapphire is proven by high-resolution scanning transmission electron microscopy (HR-STEM) and in-plane X-ray diffraction (XRD) also on a wafer-scale. Furthermore, a buried $In_2Se_3$ layer is identified via STEM-based energy dispersive X-ray spectroscopy (EDX) and X-ray reflectivity (XRR) analysis. To validate the material's electronic and structural properties, optical absorption and optical-pump terahertz-probe (OPTP) spectroscopy was performed. THz measurements indicate electron mobilities exceeding 1200 cm²/V·s — an essential parameter for high-performance logic applications.

In summary, the controllable low-temperature MOCVD process and the received mobility underline the potential for scalable and industrial integration of InSe into next-generation logic devices.

METHODS

In this work, all growth experiments were conducted using a horizontal-flow **metal-organic chemical vapor deposition** (MOCVD) reactor (*Aixtron AIX 200 GF*) equipped with gas foil rotation[69,70], a platform that has previously enabled the successful growth of various indium-based compound semiconductors including InN[71], GaInAs[72–74] or GaInNAs[75,76]. Infrared heating was used to heat the graphite susceptor. The reactor pressure was maintained at 50 mbar and hydrogen ($H_2$) was used as the carrier gas with a total flow rate of 6800 sccm for all experiments.

Moreover, we aim at precisely controlling the interface between the substrate and the 2D material, which is especially critical for the deposition of van der Waals materials, as they are – in contrast to the growth of covalently bonded crystals – grown without a buffer layer. Thereto, we introduce an In-rich seeding layer prior to the InSe growth. Prior studies on the oriented growth of 2D $WS_2$ on c-plane sapphire have shown the importance of a controlled interface reconstruction: here, a tungsten-based seeding layer was shown to promote a well-ordered interface and epitaxial alignment of the subsequent 2D layers [77]. In our own previous work on GaS, a structurally similar post-transition metal dichalcogenide[78], we also observed a gallium-rich interfacial layer at the sapphire–GaS interface[64], which played a critical role in templating the layered growth.

The synthesis was performed in a temperature range from 350 °C up to 450 °C using diisopropyl selenium (DiPSe) and tri-methyl indium (TMIn) as selenium and indium precursors, respectively. The DiPSe partial pressure was between 930 to 4060 times larger than the TMIn partial pressures. The ratio of the supplied precursors significantly exceeds the targeted incorporation ratio, which can be attributed to surface-reaction-limited growth kinetics. This behavior is well known from ZnSe synthesis using DiPSe below 440 °C[52]. Prior to growth, the c-plane sapphire substrate was preheated at 600 °C for 22 minutes in order to reduce hydroxyl (–OH) group coverage on the surface and thereby improve surface cleanliness and reproducibility of the growth conditions[79]. After lowering the temperature to 400°C, the indium precursor was introduced for 170 seconds (TMIn predeposition, In seeding step) to promote the interaction between indium and the sapphire surface, thereby facilitating the catalytic decomposition of the selenium precursor introduced during the growth phase. This step ensures that selenium incorporation occurs from the onset of growth. After the growth time $t_G$ the precursor flow was stopped and the reactor was cooled to room temperature.

**Atomic force microscopy** (AFM) was used to analyze the sample topography using a Nanosurf Flex AFM system operated in dynamic force mode (tapping mode). Gwyddion software was used for data analysis and processing. Commercial probes from BudgetSensors Innovative Solutions Bulgaria Ltd. with a nominal tip radius of 10 nm were used for AFM measurements.

**Optical Absorption spectra** were acquired in transmission geometry (spot size ~ 3 mm diameter) using an OceanOptics HDX-XR spectrometer (spectral resolution Δλ = 1.1 nm) together with an OceanOptics DH-2000 combined halogen and deuterium light source. This enables both illumination and detection of light in the range from 205 nm – 1100 nm (1.13 eV – 6 eV).

**Terahertz (THz) transmissions spectroscopy:** The OPTP setup utilizes a 5 kHz regenerative amplifier (Spectra Physics Solstice Ace) delivering 50 fs pulses spectrally centered around 800 nm. The output is split into three parts. The first part excites a large-aperture LT-grown GaAs antenna, which generates ∼1 ps long terahertz (THz) pulses. These pulses are then transmitted through the InSe samples to probe their photoinduced conductivity. The second part serves as the gate pulse for electro-optic sampling, using a 500 μm thick ZnTe crystal cut along the ⟨110⟩ orientation. This enables the measurement of the THz electric field in the time domain. The third part of the 800 nm output is directed into an optical parametric amplifier (OPA, Spectra Physics TOPAS) to generate pump pulses with a central photon energy of 2.6 eV (480 nm). The pump pulse is temporally delayed relative to the THz probe by a motorized linear translation stage and modulated by an optical chopper before exciting the InSe sample. A schematic illustration of the experimental setup is shown in Figure SI6.

The setup enables the detection of THz pulses covering a bandwidth from 0.3 THz to 2.3 THz, which are resolved in the time domain by electro-optic sampling. We record a time window of 9 ps, to which we apply a Blackman–Nuttall window with a 2 ps slope prior to Fourier transformation. This procedure yields the complex reference spectrum $E(\omega)$ as well as the pump-induced change $\Delta E(\omega)$ caused by the modulated pump excitation.

The corresponding change in optical conductivity $\Delta\sigma(\omega)$ is then calculated as:

$$\Delta\sigma(\omega) = (2\, c_0\, \varepsilon_0\, \sqrt{\varepsilon_r})\, /\, L\, \cdot\, (\Delta E(\omega)\, /\, E(\omega))$$

where $c_0$ is the speed of light in vacuum, $\varepsilon_0$ is the vacuum permittivity, $\varepsilon_r$ is the relative dielectric constant of the sample, and L is the effective thickness of the photoexcited layer. The resulting frequency-dependent photoconductivity is analyzed by fitting the Drude-Smith model, which is given by:

$$\sigma(\omega) = [n\, e^2\, \tau\, /\, m*] \times [1\, /\, (1 - i\, \omega\, \tau)] \times [1 + c_1\, /\, (1 - i\, \omega\, \tau)]$$

where n is the carrier density, e is the elementary charge, τ is the carrier scattering time, m* is the effective mass, and $c_1$ is the persistence of velocity parameter describing carrier backscattering.

**Raman spectroscopy** was performed in backscattering geometry using a confocal Raman microscope (Horiba XPloRA Plus) with a 100×/0.8 NA objective and an excitation wavelength of 532 nm. After averaging Raman spectra collected over at least 10 × 10 spots with a spatial step size of ≥ 2 μm, background subtraction was performed using a asymmetrically reweighted penalized least squares smoothing[80]. To identify the present $In_xSe_y$ phases, the spectra were fitted using a pseudo-Voigt function. The Lorentzian half-width at half-maximum (γ) was set to 5 cm⁻¹, with allowed variation between 4 and 6 cm⁻¹. The mixing parameter η, which defines the weighting between Gaussian (η = 0) and Lorentzian (η = 1) contributions, was fixed at 0.5 but permitted to vary within this full range to optimize the fit. Peak positions were initialized based on literature values for InSe[34], $β-In_2Se_3$[34], $γ-In_2Se_3$[34], amorphous Selenide[60] and

$In_4Se_3$[61], allowing for a deviation of ±3 cm⁻¹ to accommodate experimental and material-related factors.

**X-ray diffraction** (XRD) was used to characterize the crystalline structure and identify film phases using a Bruker D8 Discovery diffractometer with monochromated Cu-Kα radiation and a LynxEye silicon strip detector.

**X-ray reflectivity** (XRR) measurements were performed using a Panalytical X'Pert Pro diffractometer using Cu-Kα radiation. The reflectivity data were analyzed using X-ray Calc3[65] software to extract the thickness and density of the InSe layer.

**TEM Sample Preparation**: The TEM lamella was prepared using a Thermofisher Scientific Helios 5 Hydra CX, an analytical dual-beam SEM / Plasma Focused Ion Beam (PFIB) instrument. Before FIB preparation, the sample was sputter coated with carbon prior to loading into the SEM. The initial pre-coating was necessary to prevent the InSe layer from amorphization under electron beam exposure and to overcome charging issues since the substrate was sapphire (cf. Figure SI5). The selected region of the sample was additionally protected with ≈ 1 µm thin carbon and tungsten layers, respectively. This procedure ensures sufficient surface protection from the xenon ion milling and subsequent fine polishing with argon ions.

**STEM Measurement: Energy dispersive X-ray** (EDX) analysis and **high-resolution HAADF STEM** imaging were performed in a double Cs-corrected JEOL JEM-2200FS operated at 200 kV acceleration voltage equipped with a Bruker Flash 5060 EDX detector. The quantitative elemental evaluation is based on the simulated Cliff-Lorimer Methods[81] implemented in the Bruker Esprit 2.3 software. Rapid amorphization of the InSe layer was observed under the electron beam. To image the initial structure of the InSe layer, the imaging conditions were modified by reducing the sampling and dwell time per pixel of the STEM image. This resulted in a reduction of the electron dose by almost one order of magnitude from $2.8*10^5$ e/A² to $4.9*10^4$ e/A², which significantly improved the imaging of the InSe layer.

ACKNOWLEDGEMENT
The authors gratefully acknowledge support from the Deutsche Forschungsgemeinschaft (DFG) via SFB 1083 (Project number 223848855). Funding from the European Regional Development Fund (ERDF) and the Recovery Assistance for Cohesion and the Territories of Europe (REACT-EU) is gratefully acknowledged. We also thank Dockweiler Chemicals GmbH, Marburg, for collaboration regarding precursor development and purification. SC acknowledges funding through the LOEWE Transfer professorship „high-performance materials ".